\documentclass[a4,12pt]{article}
\usepackage{graphicx}
\graphicspath{}
\usepackage[numbers,sort&compress]{natbib}
\usepackage{amsmath}

\begin{document}
\title{Strongly confined 2D parabolic quantum dot: Biexciton or Quadron?}
\author{Nguyen Hong Quang$^{1,2,\footnote{Corresponding author. Email: nhquang@iop.vast.ac.vn}}$, Nguyen Que Huong$^{3}$,\\ Tran Anh Dung$^{4}$, Nguyen Toan Thang$^{1}$, and Hoang Anh Tuan$^{1}$}


\maketitle

\noindent $^1$ Institute of Physics, Vietnam Academy of Science and Technology (IOP, VAST), 18 Hoang Quoc Viet, Nghia Do, Cau Giay, Hanoi, Vietnam\\[0.2cm]
$^2$ Graduate University of Science and Technology, VAST, 18 Hoang Quoc Viet, Nghia Do, Cau Giay, Hanoi, Vietnam\\[0.2cm]
$^3$ Marshall University, One John Marshall Drive, Huntington WV 25701\\[0.2cm]
$^4$ Institute for Scientific Information, VAST, 18 Hoang Quoc Viet, Nghia Do, Cau Giay, Hanoi, Vietnam
\begin{abstract}
Excitonic systems localized in a single InAs/GaAs parabolic quantum dot are studied theoretically using an unrestricted Hartree-Fock method. The binding energies of excitons, conventional biexcitons and quadrons - four particles system of two electrons and two holes, in the ground state have been obtained as functions of magnetic field and confinement potential. It is found that for strong confinement, while the binding energy of biexciton is negative, the binding energy of quadron is positive, suggesting the strong lateral confinement of the parabolic quantum dot supports the formation of a quadron rather than a biexciton.
\end{abstract}

\noindent {\bf keywords}: 
exciton, biexciton, quadron, binding energy, parabolic quantum dot, Hartree-Fock method

\maketitle

\section{Introduction}
With recent fast development of nanotechnology, physics of semiconductors  actually becomes the physics of low-dimensional semiconductor systems. Quantum confinement of nanostructures brings unique and new physics that have never been observed in traditional semiconductors and makes the nanostructures interesting objects of intensive studies (for reviews, see Refs. \cite{Bimberg,Reimann,Wang,Keze} and cited therein references).  

Excitons and exciton-related excitations, such as biexcitons, play a very important role in optical processes happening at or close to the band gap such as single-photon emissions. Due to both their basic physics nature and their promising applications in optical quantum information technology such as quantum cryptography and quantum computing, they have attracted an enormous amount of research for various low-dimensional semiconductor systems \cite{Thao,Long,Tsuchiya,Varga,Donck,Keze2,Xie,Singh,Sujanah,Pokut1,Pokut2-4,Pokut5-6,Masumoto,Rodt,Rodt2,Ikezawa,Tsai,Schliwa,Zielinski,Quang1,Quang,Natori1,Natori,War,War2}.
Thus, the binding energies of biexcitons as a bound state of two separate excitons, were calculated in CuCl/NaCl and GaAs spherical quantum dots by Monte-Carlo method \cite{Tsuchiya}, in 2D systems by stochastic variational method \cite{Varga,Donck} and by hyperspherical harmonics \cite{Keze2}. The theory of biexcitons formed by spatially separated electrons and holes have been studied by variational method in different nanosystems, such as ZnSe quantum dots in a borosilicate glassy matrix \cite{Pokut1}, CdS, ZnSe and Al$_2$O$_3$ quantum dots in a dielectric matrix \cite{Pokut2-4}, and in Ge/Si heterostructure with germanium quantum dots \cite{Pokut5-6}. It is shown that in these nanosystems, biexciton formation is possible and of the threshold character when the spacing between the quantum dots surfaces is larger than a certain critical spacing.

Of particular interest are semiconductor quantum dots with parabolic confinement potential, including self-assembled semiconductor quantum dots \cite{Ikezawa,Schliwa,Masumoto,Rodt,Rodt2,Tsai,Zielinski,Quang1,Quang,Natori1,Natori,War,War2}, because this kind of quantum dots shows large binding energies of biexctions and, on the other hand, it has easily been incorporated in field-effect structures to study the influence of the external magnetic field while not breaking the system symmetry. It is expected that binding energies of excitons, biexciton are increased due to the spatial confinement, which reduces the exciton Bohr radius, and depend strongly on parameters of the quantum dot system such as the nanostructure materials, size, shape, spatial confinement and applied magnetic field.
The binding energies of biexcitons have been studied both experimentally and theoretically for a large variety of quantum dots with different geometry, shape and composition \cite{Ikezawa,Schliwa,Masumoto,Rodt,Rodt2,Tsai,Zielinski,Quang1}. The biexciton binding energy has been reported both positive and negative in small quantum dots, depending on the parameters of the system such as confinement, magnetic field, and other relationship of Coulomb interactions between electrons and holes \cite{Masumoto,Rodt,Rodt2}. The experimental data and theoretical calculations for biexciton binding energy in small self-assembled InAs/GaAs quantum dots have shown negative values \cite{Tsai,Zielinski} and it is a very sensitive function of the lattice randomess. 

With all the studies up to now, however, there still exists an unclear picture about how these few electrons and holes interact with each other inside the quantum dot. To our knowledge, studying many-electron-hole systems in semiconductors nanostructures, almost all works done so far rely on exciton and biexciton formations and their interaction. That means, the biexcitons are assumed to be a priori formed in these many-electron-hole systems. However, while in other systems excitons have always been considered to be formed first and later the coupling of two excitons forms the biexciton, there are some evidences showing that in quantum dots with parabolic confinement potential the formation of a new state of two electrons and two holes with equal roles and pair interaction between each others - we will call this state quadron- is also possible, and in some cases even more preferable. In fact, it is still unclear if the lateral confinement of the quantum dot supports the formation of conventional biexcitons as a bound state of two separate excitons or quadron, the four-particle excitations consisting of two electron and two holes of equal interaction.

In this work, we try to answer this question by calculating the binding energies of biexcitons and quadrons in strongly confined 2D parabolic quantum dots using the unrestricted Hartree-Fock method. We are able to show that the binding energy of biexciton in the ground state is always negative in all ranges of parameters, indicating that in its ground state biexciton is antibinding. Our calculations agree very well with the  experimental data and are consistent with the  results  calculated by other authors \cite{Rodt,Tsai,Zielinski}. Furthermore, our results demonstrate that while the binding energy of biexciton as a state of two separate excitons is always negative, the binding energy of quadron as a four-particle  states of two electrons and two holes all together is positive, and both are very sensitive to the electron-to-hole ratio of the confinement potentials. These results suggest that the strong lateral confinement of the parabolic quantum dot might support the formation of a quadron rather than a biexciton, and this point has never been noted so far.

\section{Theoretical method and model}

Many different methods have been used to study excitons and biexcitons, such as few-body system method \cite{Keze}, diffusion Monte Carlo methods \cite{Tsuchiya}, variational method \cite{Varga,Donck,Pokut1,Pokut2-4,Pokut5-6}, hyperspherical harmonics \cite{Keze2}, configuration interaction method \cite{Ikezawa,Schliwa,Rodt,Rodt2,Tsai,Zielinski}, Hartree-Fock method \cite{Quang1,Zielinski,Quang,Natori1,Natori} and perturbation theory \cite{,War,War2}.
Among these methods, the unrestricted Hatree-Fock method which seems to be a good approximation for many-electron systems in parabolic quantum dots \cite{F}, and spherical quantum dots \cite{BS}, is especially appropriate for small self-assembled quantum dots. This method has been used successfully in our previous works \cite{Quang,Natori1,Natori} to study the charging effects, as well as external magnetic field effects on charged magneto-excitons in self-assembled quantum dots. Our results agree very well with experimental results \cite{War,War2} for the energy red-shift and exciton absorption spectra in small InAs/GaAs self-assembled quantum dots.
In this work we use unrestricted Hatree-Fock method to study the exciton, biexciton and quadron in small self-assembled quantum dots with parabolic confinement potential in magnetic field.

We consider excitonic systems as a system of interacting electrons and holes confined in a 2D quantum dot with parabolic lateral potential in the presence of a perpendicular magnetic field $\vec{B} \| z$. In the framework of the effective-mass approximation, the total Hamiltonian of the system of $N$ electrons and $M$ holes (for exciton $N=M=1$, and for biexciton and quadron $N=M=2$)  with the full many-body approach can be written as follows:
\begin{eqnarray}
\widehat{H} = \sum_{i=1}^{N}h(\vec{r}_i) +\sum_{k=1}^{M}h'(\vec{r}_k) \ \ + \quad \hspace{3cm} \ && 
\nonumber \\
\ + \ \frac{1}{2}\sum_{i,j=1; i\ne j}^{N}
\frac{e^2}{\epsilon r_{ij}} +\frac{1}{2}\sum_{k,l=1; k\ne l}^{M}\frac{e^2}{\epsilon r_{kl}}-\sum_{i=1}^{N}\sum_{k=1}^M\frac{e^2}{\epsilon r_{ik}} \ , &&
\end{eqnarray}
where the first two terms are Hamiltonians of single electrons $h(\vec{r}_i)$  and single holes  $h'(\vec{r}_k)$, and the last three  terms are the total electron-electron, hole-hole and electron-hole Coulomb interactions, respectively, with $\epsilon$ being the material dielectric constant. 

The single-particle Hamiltonians for electron (with notations $e, i$) and hole (with notations $h, k$) in a quantum dot with parabolic confinement in a magnetic field are written as follows ($\hbar=c=1$):
\begin{eqnarray}
h(\vec{r}_i) = -\frac{\nabla^2_i}{2 m^*_e}
+\frac{m^*_e}{2}(\omega^2_e+\frac{1}{4}\omega^2_{ce})r^2_i
+\frac{1}{2}\omega_{ce} \hat{L}_{zi} \ ,\label{eq:He} &&\\[0.3cm]
h'(\vec{r}_k) = -\frac{\nabla^2_k}{2 m^*_h}
+\frac{m^*_h}{2}(\omega^2_h+\frac{1}{4}\omega^2_{ch})r^2_k
+\frac{1}{2}\omega_{ch} \hat{L}_{zk} \ ,\label{eq:Hh} &&
\end{eqnarray}
where  $m^*_e$ ($m^*_h$) and $\omega_e$ ($\omega_h$) are the effective mass and the confinement potential of the electron (hole), respectively; $ \omega_{ce} = eB/m^*_e$ ($ \omega_{ch} = eB/m^*_h$) and $\hat{L}_{zi}$ ($\hat{L}_{zk}$) are the cyclotron frequency and the z-components of orbital angular momentum operators of the electron (hole), respectively.  Note that as in  \cite{Quang1,Quang,Natori1, Natori} the terms describing the spin Zeeman splitting due to interaction of the spin with the magnetic field in  (\ref{eq:He}) and (\ref{eq:Hh}) has been neglected because of its smallness.

We note that in this calculation we only consider the system in the ground state, when the two electrons (as well as the two holes) have opposite spins and therefore there is no exchange interaction involved.

In the framework of the unrestricted Hartree-Fock approximation, the total wave function of the system of $N$ electrons and $M$ holes can be found in the form of direct product of the Slater determinants for electrons and holes: 
\begin{equation}
\Psi(\xi_1,...,\xi_{N},\xi'_1,...,\xi'_{M}) = |\psi_1(\xi_1)...\psi_{N}(\xi_{N})| . |\psi'_1(\xi'_1)...\psi'_{M}(\xi'_{M})| \ , 
\end{equation}
where the electron and hole orbitals $\psi_i(\xi)$, $\psi'_k(\xi)$ in the Slater determinants are spin dependent: $\psi_i(\xi)=\phi_i^\alpha (\vec r)\sigma(\alpha)$ or  $\psi_i(\xi)=\phi_i^\beta (\vec r) \sigma(\beta)$ for spin-up or spin-down electrons, and $\psi'_k(\xi)=\phi'^{\ \alpha}_k (\vec r) \sigma(\alpha)$ or  $\psi'_k(\xi)=\phi'^{\ \beta}_k (\vec r) \sigma(\beta)$ for spin-up or spin-down holes.

In the Hartree-Fock method with the Roothaan formulation \cite{Quang1,Quang,Natori1,Natori}, the spatial parts of electron and hole orbitals $\phi_i^{\alpha, \beta}(\vec r)$ \ and  \ $\phi'^{\ \alpha, \beta}_k(\vec r)$ are written in the form of expansions in the basis functions $\chi^e_{\nu}(\vec{r})$ and $\chi^h_{\mu}(\vec{r})$, which are chosen as the eigen-functions of the single particle Hamiltonians (\ref{eq:He}) and (\ref{eq:Hh}):
\begin{eqnarray}
\phi_i^{\alpha, \beta}(\vec r) &=& \sum_{\nu}C_{i \nu}^{\alpha, \beta}\chi^e_\nu(\vec r) \ , \\
\phi_k'^{\ \alpha, \beta}(\vec r) &=& \sum_{\mu}C_{k \mu}'^{\ \alpha, \beta}\chi^h_\mu(\vec r) \ ,
 \end{eqnarray}
 where indexes - quantum numbers $\nu,\mu \equiv (n,m)$ run over all single electron or hole states described by the wave-functions in polar coordinates $\chi^e_{\nu}(\vec{r})$ and $\chi^h_{\mu}(\vec{r})$ :
\begin{eqnarray}
\label{eq:chi}
\chi^{e,h}_{n,m}(r,\varphi)=\frac{1}{\sqrt{2\pi}}e^{i m \varphi}\sqrt{\frac{2 n!}{(n+|m|)!}}\  \times \hspace{2cm} && \nonumber \\  \times \ \alpha_{e,h}(\alpha_{e,h} r)^{|m|} e^{-(\alpha_{e,h} r)^2/2} L_n^{|m|}((\alpha_{e,h} r)^2), &&
\end{eqnarray}
where $L_n^{|m|}(r)$ is generalized Laguerre polynomial, and 
\begin{eqnarray}
\alpha_{e,h} &=& \sqrt{m_{e,h}^* \Omega_{e,h}} \ , \nonumber\\
\Omega_e &=& (\omega_e^2+\frac{1}{4}\omega_{c_e}^2)^{1/2} \ , \\
 \Omega_h &=& (\omega_h^2+\frac{1}{4}\omega_{c_h}^2)^{1/2} \ . \nonumber
\end{eqnarray}

Note that in the chosen basis (\ref{eq:chi}), the electron-electron, hole-hole and electron-hole Coulomb interaction matrix elements in Hartree-Fock Roothaan equations can be calculated analytically (see {\it e.g.} \cite{Halo}). We solve self-consistently the unrestricted Hartree-Fock Roothaan equations to obtain the total energy of the system for exciton ($E_{X}$), biexciton ($E_{XX}$), and quadron ($E_{Q}$) in the dependence on magnetic field and confinement. Then the binding energies of the excitonic systems are found by the following equations:
\begin{eqnarray}\label{eq:bin}
E^{b}_{X}&=&(E^e+E^h)-E_{X} , \  \text{for exciton} , \nonumber\\ 
E^{b}_{XX}&=&(E_X +E_X)-E_{XX} , \ \text{for biexciton} , \\
E^b_{Q}&=&(2E^e+2E^h)-E_{Q} ,  \ \text{for quadron}.\nonumber 
\end{eqnarray}

\section{Numerical results and discussions}

In this section, in order to compare with the experimental results obtained in \cite{Tsai} for biexciton binding energy, we will present the numerical results using the parameters appropriate for InAs/GaAs self-assembled quantum dots: $m^*_e = 0.067 m_o$, $m^*_h = 0.25 m_o$,  $\omega_e = 49$ meV, $\omega_h = 25$ meV,$ \ \epsilon = 12.53$ \cite{Natori, Quang}. The effective Bohr radius $a^*_B = \epsilon/m^*_e e^2=9.9$ nm has been adopted as an unit of length and two times of the effective Rydberg $2Ry^* = m^*_e e^4/\epsilon^2=11.61$ meV as an unit of energy. 
For small InAs/GaAs self-assembled quantum dots, the oscillator lengths for electrons and holes in the absence of magnetic fields $l_{e, h} = (m^*_{e, h} \omega_{e, h})^{-1/2}$ are $4.8$\ nm and $3.5$\ nm, respectively. These values are much smaller than the effective excitonic Bohr radius which is about $13$\ nm, what means  electrons and holes in small InAs/GaAs dots are strongly confined.

In the investigation of the impact of magnetic field, for comparison purpose, beside the above-mentioned parameters (called parameter set 1), in order to see the sensitive changes of binding energy depending on related parameters, the numerical calculation using another parameter set (called parameter set 2), with $m^*_e = 0.067 m_o$,\ $m^*_h = 0.067 m_o$,\ $\omega_e = 49$ meV,\  $\omega_h = 25$ meV, has also been performed.

The dependence of the biexciton binding energy on the magnetic field  has been studied and presented in Fig.1, where the calculations performed for two parameters sets have been compared. The results show that for both parameter sets the biexciton binding energy are increased with the magnetic field, as expected, due to the additional confinement the magnetic field applies on the electron and hole, however the increase is rather small. 

It is important to note that for all variation range of magnetic field for both parameter sets, the biexciton binding energy in ground state is always negative. The difference of binding energies calculated by two sets is about 2 meV. The result shows that the biexciton in ground state is antibinding and unstable in these systems. 
\begin{figure}[htbp]
\centering \includegraphics[width=6cm]{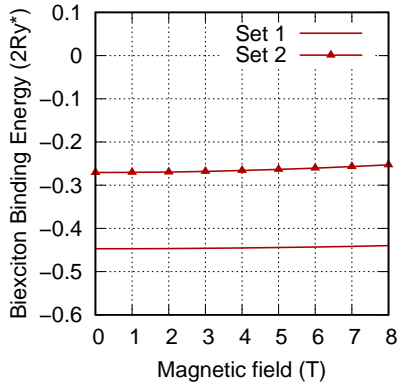}
\caption{\label{Fig1}(color online) The binding energy of biexciton as function of magnetic fields for two sets of parameters (see in the text).}
\end{figure}
Our calculation results agree well with the experimental results \cite{Tsai}. Indeed, from the calculation for parameter set 1 and set 2, the biexciton binding energy is $-0.45 (2Ry^*)$ and $-0.25 (2Ry^*)$ (see Fig.\ref{Fig1}), (or $-5.1$ meV and $-2.9$ meV), respectively, which is in very good agreement with the range from $-6$ meV to $-1$ meV of experimental data for different dots \cite{Tsai}. Note that these results also agree with the negative biexciton binding energies calculated by the configuration interaction method in \cite{Rodt} and  \cite{Zielinski}.  The antibinding property of an biexciton is a consequence of the strong confining potential, which assists direct Coulomb repulsive interaction between excitons. But our study shows it is not the case for the quadtron - the four-particle excitation of two electrons and two holes.
\begin{figure}[htbp]
\centering
\includegraphics[width=6cm]{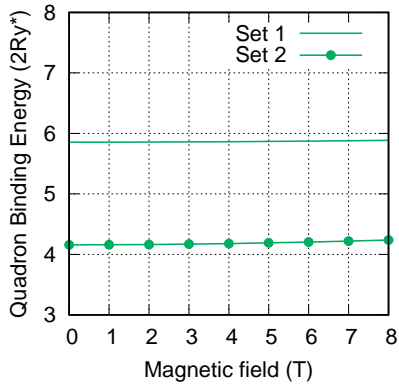}
\caption{\label{Fig2}(color online) The binding energy of quadron as function of magnetic fields for two sets of parameters (see in the text).}
\end{figure}

In order to understand the nature of the interactions in the quatron, the binding energy of quadron - the four-particle excitation of two electrons and two holes with the full many-body approach \cite{Zielinski} when the interactions between all pairs of particles have been taken into account equally- has been calculated.  In Fig. 2 the binding energies of a quadron in the ground state as a function of magnetic field for the above-mentioned two sets of parameters is presented. In contrast to biexciton case, the quadron binding energy is always positive for both parameter sets, what means the quadron is bound as a whole.

\begin{figure}[htbp]
\centering\includegraphics[width=6cm]{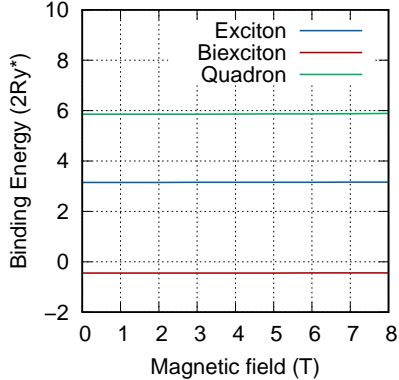}
\caption{\label{Fig3}(color online) The binding energies of exciton, quadron and biexciton as function of magnetic fields for parameters set 1 (see in the text).}
\end{figure}

To compare and see the relationship of the binding energy between three excitonic systems, in Fig.3 all the binding energies of the exciton, biexciton and quadron have been shown for the case of parameter set 1. As seen from Fig.3, the impact of a magnetic field on the binding energies in the strong confinement is rather small, but it is very important to notice that while the biexciton binding energy is negative, the quadron binding energy is positive. Since the energy of the system of four particles is the same in our calculation, $E_Q=E_{XX}$, the equation describing the relationship between the binding energies of the quadron, exciton, and biexciton defined by (\ref{eq:bin}), could be written as follows:
\begin{equation}\label{eq:rel}
   E^b_{Q} =  2 E^b_{X} + E^b_{XX}. 
\end{equation}  

Self-assembled quantum dots, although relatively homogeneous, in practice are still different in size resulting in different confinements for electron and holes. Therefore to provide more information on the effect of random fluctuations of sizes, we investigate the dependence of binding energies on the various confinement values via the ratio of confinement potential between electrons and holes $\omega_e/\omega_h$.
Thus, the binding enegies of exciton, biexciton and quadron have been calculated for a wider range of the electron-to-hole lateral confinement ratio $\omega_e/\omega_h$ with two sets of parameters:  ($\omega_e=49 \text{ meV}, m_e^*=0.067, m_h^*=0.25$) as in parameter set 1 and  ($\omega_e=49 \text{ meV}, m_e^*=0.067, m_h^*=0.067$) as in parameter set 2, respectively, but with $\omega_h$ changing.

Fig.4 and Fig.5 show the binding energy of the exciton, biexciton, and quadron in the absence of magnetic field for the parameter set 1 and set 2, respectively.  As seen from Fig.4 and Fig.5, the binding energy of biexciton is always negative while the binding energy of quadron is positive. One can also see that the relationship between the binding energies of exciton, biexciton and quadrons defined by (\ref{eq:rel}) holds exactly for all range of magnetic field and confinement potential in the present study (see Fig.3, Fig.4 and Fig.5). 
\begin{figure}[htbp]
\centering\includegraphics[width=6cm]{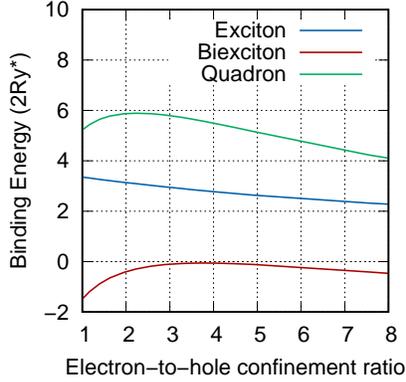}
\caption{\label{Fig4}(color online) The binding energy of exciton, biexciton, and quadron as function of ratio $\omega_e/\omega_h$ with  $B=0$ T,  $\omega_e=49 \text{ meV}, m_e^*=0.067, m_h^*=0.25$.}
\end{figure}
\begin{figure}[htbp]
\centering \includegraphics[width=6cm]{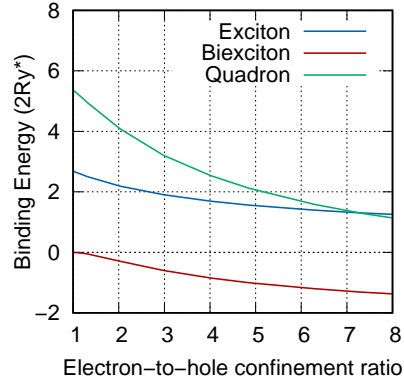}
\caption{\label{Fig5}(color online) The binding energy of exciton, biexciton, and quadron as function of ratio $\omega_e/\omega_h$ with $B=0$ T, $\omega_e=49 \text{ meV}, m_e^*= m_h^* = 0.067$.}
\end{figure}

Comparing the results obtained for two parameter sets in Fig.4 and Fig.5, one can see that the binding energies of exciton, biexciton and quadron are very sensitive to the system parameters. This can be understood qualitatively as the variation of mass and confinement ratio between electron and hole results in the changes of the effective lengths of the electron and hole. As a consequence, this changes magnitudes and correlations between the Coulomb interactions of electron-electron, hole-hole, and electron-hole pairs, which result in changes of the binding energies. These results are indicative for studies of quantum dots with different material parameters, and can be verified by experiments.

\begin{figure}[htbp]
\centering \includegraphics[width=6cm]{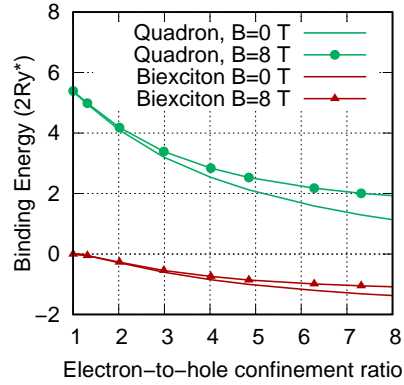}
\caption{\label{Fig6}(color online) The quadron and biexciton binding energy as function of ratio $\omega_e/\omega_h$ with $\omega_e=49 \text{ meV}, m_e^*=m_h^*=0.067$, at $B=0$ T and $B=8$ T.}
\end{figure}

To see the influence of the magnetic field on the binding energies, in Fig.6 the biexciton and quadron binding energies are compared for two cases with and without magnetic field: $B=0$ T and $B=8$ T. One can see that the magnetic field plays a significant role at a large difference in confinements of electron and hole, or in other words, large difference in effective lengths of the electron and hole. For all ranges of calculation in our present study, while the binding of biexciton is negative,  the binding of quadron is positive. This very interesting result leads us to a serious conclusion that  the strong lateral confinement in parabolic quantum dot tends to support the equal multi-interaction between particles and the formation of the quadron, rather than biexciton. The question of preferability of the system  to form either biexciton or quadron in the quantum dots has never been raised in literature so far and needs more study.

\section{Conclusion}

In conclusion, in this paper the properties of excitons, biexcitons and quadrons in small InAs/GaAs self-assembled quantum dots with strong lateral parabolic  confinement have been studied. Our theoretical calculations by using Hartree-Fock-Roothan method show that the biexciton binding energy in the ground states is always negative for all variation range of magnetic field and of quantum dots parameters, indicating that the biexciton in the ground state is antibinding, which agree very well with the experimental data and the previous results by other authors. 
The dependence of the binding energies as functions of electron-to-hole confinement ratio for a wider range of quantum dots has been showed. The results can be useful for experimental checks for various quantum dots with different material parameters. 
Finally, we introduced the first time quadron, the new excitation of two electrons and two holes with their pair interaction taken into account equally. Per our results while the binding energy of biexciton is always negative, the binding energy of quadron is positive, and both are very sensitive to the quantum dots parameters. The relationship between the binding energies of the exciton, biexciton and quadron has been established. Our results suggest that the strong lateral confinement of the parabolic quantum dot preferably supports the formation of a quadron rather than a biexciton. These findings might serve as a suggestion for experimental detection of quadrons in quantum dots.

\section*{Acknowledgments}

This work is supported by the International Centre of Physics (ICP), Institute of Physics, Vietnam Academy of Science and Technology (VAST) under project No. ICP.2019.01. The authors thank Profs Akiko Natori, Philippe Dollfus,  Nguyen Ba An and Nguyen Nhu Dat for their careful reading of the manuscript and helpful remarks.

\end{document}